\pdfoutput=1

\documentclass[11pt]{article}

\usepackage[final]{acl}

\usepackage{times}
\usepackage{latexsym}

\usepackage[T1]{fontenc}

\usepackage[utf8]{inputenc}

\usepackage{microtype}

\usepackage{inconsolata}

\usepackage{graphicx}

\usepackage{amsmath}
\usepackage{listings}
\usepackage{makecell}
\usepackage{multirow}
\usepackage{tabularx}

\title{Observing the Southern US Culture of Honor Using Large-Scale Social Media Analysis}

\author{Juho Kim \\
  Faculty of Applied Science and Engineering \\
  University of Toronto \\
  Toronto, Ontario, Canada \\
  \texttt{\href{mailto:juho.kim@mail.utoronto.ca}{juho.kim@mail.utoronto.ca}} \\\And
  Michael Guerzhoy \\
  Division of Engineering Science \\
  University of Toronto \\
  Toronto, Ontario, Canada \\
  \texttt{\href{mailto:guerzhoy@mie.utoronto.ca}{guerzhoy@cs.toronto.edu}} \\}

\begin{document}
\maketitle
\begin{abstract}
A \textit{culture of honor} refers to a social system where individuals' status, reputation, and esteem play a central role in governing interpersonal relations. Past works have associated this concept with the United States (US) South and related with it various traits such as higher sensitivity to insult, a higher value on reputation, and a tendency to react violently to insults. In this paper, we hypothesize and confirm that internet users from the US South, where a \textit{culture of honor} is more prevalent, are more likely to display a trait predicted by their belonging to a \textit{culture of honor}. Specifically, we test the hypothesis that US Southerners are more likely to retaliate to personal attacks by personally attacking back. We leverage OpenAI's GPT-3.5 API to both geolocate internet users and to automatically detect whether users are insulting each other. We validate the use of GPT-3.5 by measuring its performance on manually-labeled subsets of the data. Our work demonstrates the potential of formulating a hypothesis based on a conceptual framework, operationalizing it in a way that is amenable to large-scale LLM-aided analysis, manually validating the use of the LLM, and drawing a conclusion. 
\end{abstract}

\section{Introduction}

A \textit{culture of honor} refers to a social system where individuals' status, reputation, and esteem play a central role in governing interpersonal relations. In such cultures, maintaining and defending one's honor and that of one's family or group is of paramount importance. A perceived slight, insult, or challenge to one's honor often necessitates a response, which could range from verbal defense to physical retaliation, to restore the lost esteem and reputation. These cultures are sometimes theorized as emerging in societies where centralized authority is weak or absent, and where individuals must rely on their reputation and the fear of retaliation to deter aggression or mistreatment by others.

We hypothesize and confirm that internet users who belong to a region where a \textit{culture of honor} is more prevalent would be more likely to display traits predicted by their belonging to a \textit{culture of honor}. Specifically, we test the hypothesis that US Southerners are more likely to retaliate to personal attacks by attacking their attacker back.

To analyze data at scale, we leverage OpenAI's API for GPT-3.5~\cite{NEURIPS2020_1457c0d6} to both geolocate internet users and to automatically detect whether users are insulting each other.

The concept of honor has been explored in social psychology, philosophy, and literature, with authors often developing a nuanced and intricate conceptual framework. The concept has been empirically validated, notably by \citet{nisbett96}, on small samples.

However, large-scale empirical analysis of the concept of honor has been very expensive until very recently. We show that LLMs, and specifically GPT-3.5, can be used to address this. In this work, we explore a specific prediction made by the conceptual framework of the \textit{culture of honor} (and the claim that the US South is such a culture).

We emphasize connecting the conceptual framework of \textit{culture of honor} to testable empirical hypotheses. We propose further testable hypotheses arising from philosophical and descriptive work on \textit{culture of honor} that are also testable by leveraging LLMs to label data while validating the outputs of the LLMs.

\section{Related Works}

\citet{nisbett96} pioneered the concept of a \textit{culture of honor} as it pertains to the US South, identifying traits such as a higher sensitivity to insult, a higher value on reputation, and a tendency to react violently to insults.

To further support their hypothesis, \citet{nisbett96} have conducted both observational and empirical studies on groups of Northern and Southern college students. In a series of experiments, the subjects were put into a situation where one would feel ``diminished.'' Then, the researchers examined whether the subjects would ``take [an] aggressive action to compensate for the diminishment they experience.'' The results conclusively demonstrated that people in the Southern US tend to exhibit the traits associated with \textit{cultures of honor} than their Northern counterparts. Taking inspiration from their previous work, our study extends  \citet{nisbett96}'s work in the online domain.

The concept of honor has been explored since ancient times, e.g., in the Bible and by Sophocles in, e.g., \textit{Antigone}\footnote{We distinguish between exploring and conceptualizing honor from referring to the psychological and cultural construct. Other ancient sources that do not necessarily conceptualize and explore honor but do refer to it extensively include Homer and Confucius.}. More recently, \citet{appiah2011honor} and \citet{sommers2018honor} connected the ancient concept, \citet{nisbett96}'s work, and modern philosophy. We draw on more recent philosophical work to make concrete predictions about internet denizens who belong to \textit{cultures of honor}.

\subsection{How a Culture of Honor Might Manifest on the Internet}

In a \textit{culture of honor}, a perceived attack or slight calls for retaliation to defend one's honor and reputation~\cite{sommers2018honor, appiah2011honor}. On an internet forum where users have fixed usernames, and particularly for users whose username can be connected to other social media possibly to their real name, it is plausible that members of \textit{cultures of honor} would be more likely to retaliate against verbal attacks. We hypothesize that more members of \textit{cultures of honor} would be geolocated to the Southern US and that those geolocated to the Southern US would therefore be more likely to verbally retaliate in internet arguments.

There is prior work on detection aggression with both classical and Transformer-based methods~\cite{warner2012detecting, SADIQ2021120, Ghosh_Priyankar_Ekbal_Bhattacharyya_2023}, with the best results reporting $>90\%$ precision/recall figures. \citet{10.1162/coli_a_00502} predict future (as opposed to present) toxicity (as opposed to aggressiveness) on a different subset of the dataset we use~\cite{zhang-etal-2018-conversations}, with the best results obtained by fine-tuning RoBERTa~\cite{liu2019roberta}, at 64\% F1.

\section{Methodology}

We obtain data on internet user conversations and fine-tune GPT-3.5 models to assign users to the corresponding US regions (South, non-South, or N/A) as well as to label each part of the conversation as whether or not it constitutes a personal attack. We then analyze the data to compare the rate at which Southern and non-Southern users retaliate when personally attacked by personally attacking back.

The \texttt{subreddit-changemyview} dataset from \texttt{ConvoKit}~\cite{chang-etal-2020-convokit} was considered for our analysis. While we also analyzed other datasets in ConvoKit -- \texttt{reddit-corpus-small}, \texttt{wikiconv} \cite{hua-etal-2018-wikiconv} (from 2015 to 2018, inclusive), and \texttt{wiki-corpus} \cite{Danescu-Niculescu-Mizil+al:12a} -- we were able to obtain data of interactions involving personal attacks with sufficient statistical power in the \texttt{subreddit-changemyview} dataset only.

\begin{table*}[htp]
  \centering
  \caption{The number of speakers with whom we were able to associate a location (through the matching X profile's self-declared location) and the number of speakers with whom we were able to associate a US region (as the associated location contained enough information to discern their US region). In this table, we only show the \texttt{subreddit-changemyview} dataset since the others do not contain a sufficient number of attacks. An expanded version of this table, showing all other datasets is shown in Appendix \ref{sec:fullloc}. \label{tab:loc}}
  \centering
  \begin{tabular}{|c||c|c|c|}
    \hline
    Dataset & \# Speakers & With Location (\%) & With US Region (\%) \\
    \hline\hline
    \texttt{subreddit-changemyview} & 119889 & 7.6 & 2.9 \\
    \hline
  \end{tabular}
\end{table*}

We attempted to locate the users by fetching the self-declared locations from user profiles with the matching usernames on the social media platform X, formerly known as Twitter. Note that not all self-declared locations could be associated with a US region because some locations do not clearly delineate whether they belong to the US South or US non-South (e.g. ``Moon,'' ``United States,'' or ``Barcelona, Spain''). For the \texttt{subreddit-changemyview} dataset, the number of speakers for which we were able to associate a location and the number of speakers for which we were able to associate a US region are tabulated in Table \ref{tab:loc}. For an expanded table showing the numbers for datasets other than \texttt{subreddit-changemyview}, please consult Appendix \ref{sec:fullloc}.

In order to obtain the location-US region pairs for fine-tuning and validating our geolocation model, samples of the collected locations were manually labeled by a human annotator and divided into training and validation sets (both of size 100).

For the personal attack classifier, we used a pre-labeled dataset of personal attacks~\cite{zhang-etal-2018-conversations}. The dataset Zhang et al. was heavily unbalanced with 2094 positive and 27927 (3833 of which are section headers) negative labels. 2094 non-section header negative entries were randomly selected and merged with the positively labeled entries. Then, they were evenly divided into perfectly balanced training and validation sets.

The training sets for each task were then used to fine-tune~\cite{bommasani2021opportunities} GPT-3.5 models for the tasks of detecting personal attacks and identifying the US regions (for more details, see Appendix \ref{sec:ftjobs}). We then test the fine-tuned GPT-3.5 models by using the validation datasets. This step is essential: we empirically demonstrate that the fine-tuned models work as expected rather than blindly relying on GPT outputs.

Note that the high validation accuracies (96.0\% for personal attack detector and 100.0\% for US region classifier) show that the training set size used to fine-tune the two models is sufficient and that the fine-tuned models perform very accurately.

We are interested in testing whether   US Southerners are more likely to retaliate with personal attacks when personally attacked, compared to their non-Southern counterparts. We measure this tendency by comparing their ``retaliation'' rates. Our main hypothesis is that the users from the US South have a higher ``retaliation'' rate than their non-Southern counterparts.

We also introduce two other metrics: ``aggression'' and ``response'' rates to see if the Southerners and non-Southerners show different tendencies in other regards. The three metrics we use are defined as follows:

\begin{enumerate}
  \item Aggression: The rates of posting personal attacks.
  \item Response: The rates of responding back if personally attacked.
  \item Retaliation: The rates of personally attacking back when responding to a personal attack.
\end{enumerate}

These three rates are computed from the aforementioned conversational datasets in two different counting schemes: per user and per interaction (for users in multiple interactions, the rate is the average of 0s and 1s, with 1s indicating a positive label).

We also review some conversations containing the postings we flag as retaliatory to see if they indeed are examples of online fights in order to ensure the correctness of our analysis.

\section{Analysis}

\begin{table*}
  \caption{The per-speaker and per-interaction rates for each metric and US region. The metric labels ``AGG,'' ``RESP,'' and ``RET'' represent aggression, response, and retaliation rates, respectively. For each rate, the sample size is equal to the number of people who posted, were personally attacked, and responded to a personal attack at least once, respectively. In this table, we only show the \texttt{subreddit-changemyview} dataset since the others do not contain a sufficient number of attacks. An expanded version of this table, showing all other datasets is shown in Appendix \ref{sec:fullmetrics}. \label{tab:metrics}}
  \centering
  \begin{tabular}{|c|c|c||c|c|c|c|}
    \hline
    \multirow{2}{*}{Dataset} & \multirow{2}{*}{Per} & \multirow{2}{*}{Metric} & \multicolumn{2}{c|}{Non-South} & \multicolumn{2}{c|}{South} \\
    \cline{4-7}
    & & & Rate (\%) & \# Samples & Rate (\%) & \# Samples \\
    \hline\hline
    \multirow{6}{*}{\small \texttt{subreddit-changemyview}} & \multirow{3}{*}{Speaker} & AGG & 9.1 & 4491 & 9.4 & 1674 \\
    \cline{3-7}
    & & RESP & 51.4 & 1318 & 51.4 & 453 \\
    \cline{3-7}
    & & RET & 20.6 & 849 & 25.3 & 292 \\
    \cline{2-7}
    & \multirow{3}{*}{Interaction} & AGG & 8.8 & 8507/96146 & 9.4 & 4820/51365 \\
    \cline{3-7}
    & & RESP & 54.5 & 4113/7552 & 56.3 & 2658/4717 \\
    \cline{3-7}
    & & RET & 23.0 & 947/4113 & 28.6 & 761/2658 \\
    \hline
  \end{tabular}
\end{table*}

\begin{figure}[!htp]
\centering
\includegraphics[width=\linewidth]{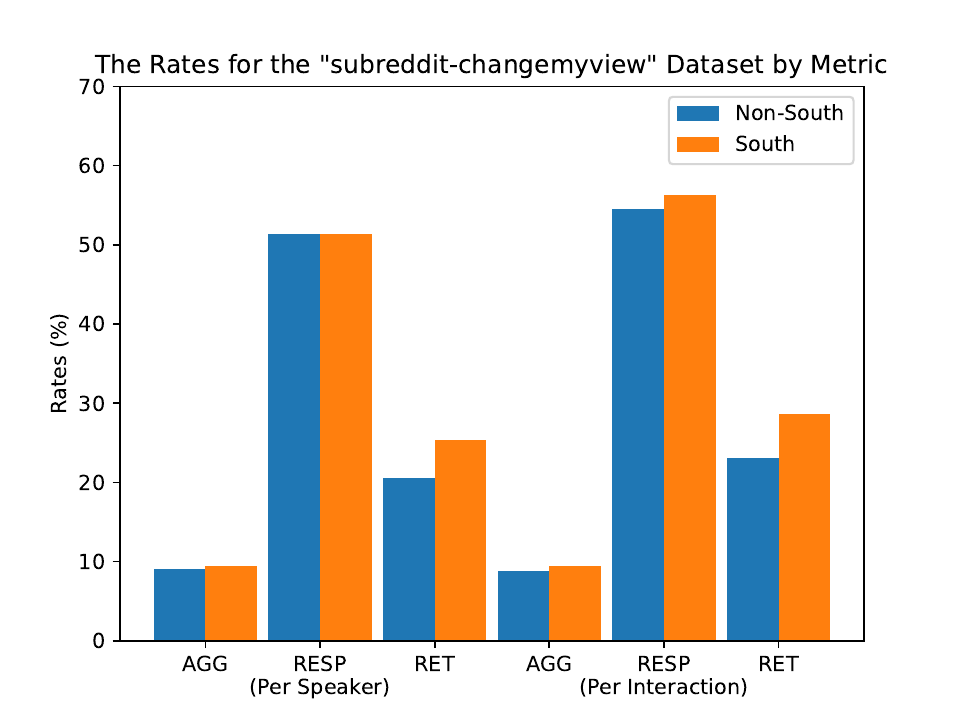}
\caption{The bar graphs of all rates for the \texttt{subreddit-changemyview} dataset by metrics and US regions. The metric labels ``AGG,'' ``RESP,'' and ``RET'' represent aggression, response, and retaliation rates, respectively. Note that the retaliation rates corresponding to the US South are greater than or equal to those of the US non-South.}
\label{fig:cmv}
\end{figure}

The computed rates for each scenario are summarized in Table \ref{tab:metrics}. Only the dataset from \texttt{subreddit-changemyview} contains samples with enough statistical power and therefore is the only one considered for further analysis. 

Note that it is not appropriate to test statistical hypotheses when a test is underpowered~\cite{button2013power}, nor is it appropriate to draw conclusions if a statistical test cannot be run. We include a summary for datasets other than \texttt{subreddit-changemyview} for completeness and transparency in Appendix \ref{sec:fullmetrics}.

For the \texttt{subreddit-changemyview} dataset, both the per-interaction and per-speaker rates for this dataset are graphed in Figure \ref{fig:cmv}. While the differences in the per-speaker rates across the US regions are marginal for aggression (+0.3\%) and non-existent for response (0.0\%), a notable difference (+5.6\%) in the retaliation rate is observed.

We fit a mixed-effect model for the per-speaker retaliation rates in \texttt{subreddit-changemyview}:

\begin{equation}
  \begin{split}
  \textrm{retaliation[i]} \sim \textrm{Bernoulli}(logit^{-1}(\beta_0 + \\ \alpha_{\textrm{regions[speakers[i]]}} + \\ \gamma_{\textrm{regions[i]}}))  \label{eq:a}
  \end{split} 
\end{equation}

\begin{equation}
  \alpha_j \sim \mathcal{N}(0, \sigma_{\alpha}) \label{eq:b}
\end{equation}

In the above Equations \ref{eq:a} and \ref{eq:b}, for interaction $i$ between two users, the original poster ``$\textrm{speakers[i]}$'' (who is also a potential retaliator) did retaliate if ``$\textrm{retaliation[i]} = 1$''. The index ``$\textrm{regions[speakers[i]]}$'' is the region (South/non-South) to which the speaker was geolocated. $\gamma_{\textrm{regions[i]}}$ is the fixed effect of regions ``$\textrm{regions[i]}$'' (the main coefficient of interest in this analysis), and $\alpha_j$ is the random effect of speaker $j$. 

Priors over parameters that are not mentioned above are flat.

The mixed-effect model estimates the speaker random effects using partial pooling, which is appropriate in our situation, where the large majority of speakers have only a few (or a single) interactions.

We find that people from the US South are more likely to retaliate (odds ratio $1.2\pm0.1$, p-value $< 0.002$) than people from outside the US South. The R code used to carry out this analysis is shown in Appendix \ref{sec:rcode}.


To further validate our analysis, 100 random samples of flagged retaliatory conversations for all datasets were reviewed to see if they were part of an ongoing online fight. For the \texttt{subreddit-changemyview} dataset which we used to prove our hypothesis, we find the actual fight percentages to be 86.0\%. Similar numbers for datasets other than \texttt{subreddit-changemyview} are observed in Appendix \ref{sec:fullfp}.

\section{Discussion}

The three landmark experiments performed by \citet{nisbett96} ``did not produce any truly violent behavior in [their] subjects, so it is an extrapolation to say that [they] have shown the process by which an insult results in actual violence for members of a \textit{culture of honor}.'' Indeed, they measured more indirect attributes of \textit{cultures of honor} such as ratings given by third-party observers and physiological changes to draw their conclusions. In contrast, our methodology demonstrated that the Southerners are more likely to signal verbal violence or capacity for verbal violence to protect their reputation in their online interactions. 

\section{Conclusion}

In this study, we were able to confirm a theoretically motivated prediction: internet users from the US South are more likely to retaliate against verbal aggression on the theory that reputation must be aggressively defended in a \textit{culture of honor}.

We show that analyses such as ours, on an extremely large scale, are possible by fine-tuning GPT-3.5 and manually verifying its outputs on samples of the inputs. By following similar processes laid out in this paper, analyses other than comparing the US South and the US non-South populations or testing hypotheses presented by social science frameworks aside from \textit{culture of honor} are also possible.

\section*{Limitations}


Our analysis is limited to the US and to English-speaking users on Reddit. \textit{Cultures of honor} as they manifest online may in principle be analyzed for other countries and languages.


The geolocation of the user may be incorrect, and, while the concentration of people who can be said to belong to a \textit{culture of honor} to some extent in the US South is thought to be higher, it is neither the case that all US Southerners are members of a \textit{culture of honor} or that there are no member of a \textit{culture of honor} elsewhere. The geolocation we obtain is merely a proxy variable. Note, however, that one in general would expect it to be more difficult to detect a trend using a proxy variable than using a direct measurement.

We validate our LLM-based analysis by checking that the LLM output is consistent with our hand-labeled sample. However, it is possible that LLMs are poorly calibrated for, for example, detecting insults in certain underrepresented dialects. Although our analysis of the hand-labeled sample means that our classifiers are correct on average, it is possible that a manual analysis would drive researchers to discover particular patterns for subgroups of the users. For example, if insults in a particular small dialect are not detectable by an LLM, it is conceivable that for speakers of that dialect, our findings would be reversed.

Our statistical analysis does not account for attenuation and error-in-variables. That is accepted practice in many analyses, where the effect of those is very difficult to estimate (and one can argue that the arbitrary 95\% significance threshold would be different if one were to upper-bound the effects of attenuation and error-in-variables).

\section*{Ethics Statement}

All research carried out in this paper was based on publicly available datasets. However, some users in the datasets may not want to be associated with \textit{cultures of honor}. We only use the geolocation of users in our analysis.

Membership in a \textit{culture of honor} can be viewed as a possibly harmful stereotype. It is important to approach broad cultural labels with nuance, humility, and respect, and to not apply broad stereotypes to individuals. However, we believe that research into \textit{cultures of honor} is appropriate: honor, as a sociological and psychological phenomenon, has fascinated humanity for millennia, and understanding \textit{cultures of honor} is important both philosophically and practically. Further,  most members of \textit{cultures of honor} are proud of their membership~\cite{sommers2018honor} and do not view affinity to that kind of culture as a harmful stereotype.

Our research only uses publically-available data and as such is not human-subjects research and is exempt from ethics board review.

The relevant subset of the datasets we used with labels from the fine-tuned models is available on Zenodo~\cite{kim_2024_10871014}.

\bibliography{./references}

\appendix

\section{Statistics on the Users with Identifiable US Region Information}
\label{sec:fullloc}

We attempted to locate the users from both Reddit and Wikipedia by fetching the self-declared locations from user profiles with the matching usernames on the social media platform X, formerly known as Twitter. In Wikipedia, anonymous editors are assigned their IP addresses as their usernames from which the associated locations can be obtained (including the associated US region). The full statistics regarding user locations for each dataset considered are shown in Table \ref{tab:fullloc}.

\begin{table*}[htp]
  \centering
  \caption{The number of speakers with an IP address as its username (relevant to Wikipedia only), the number of speakers with whom we were able to associate a location (either through the matching X profile's self-declared location or their IP address), and the number of speakers with whom we were able to associate a US region (as the associated location contained enough information to discern their US region) for each dataset. \label{tab:fullloc}}
  \centering
  \begin{tabular}{|c||c|c|c|c|}
    \hline
    Dataset & \# Speakers & With IP (\%) & With Location (\%) & With US Region (\%) \\
    \hline\hline
    {\small \textbf{\texttt{subreddit-changemyview}}} & \textbf{119889} & \textbf{--} & \textbf{7.6} & \textbf{2.9} \\
    \hline
    \texttt{reddit-corpus-small} & 217100 & -- & 7.5 & 2.8 \\
    \hline
    \texttt{wikiconv} & 621142 & 39.1 & 40.9 & 16.7 \\
    \hline
    \texttt{wiki-corpus} & 38462 & 7.8 & 11.4 & 4.8 \\
    \hline
  \end{tabular}
\end{table*}

\section{Information on the Fine-Tuned GPT-3.5 Models}
\label{sec:ftjobs}

Tables \ref{tab:usrc} and \ref{tab:pad} describe information relevant to the fine-tuning of the GPT-3.5 models for the tasks of US region classification and personal attack detection, respectively.

\begin{table*}[htp]
  \centering
  \caption{The system prompts, default classifications (in case the model produces an unexpected output), accuracies, and training/validation information of the fine-tuned GPT-3.5 model for the task of US region classification. Note that the system prompt was broken to fit it in the page. The actual prompt is single-line. \label{tab:usrc}}
  \begin{tabularx}{\textwidth}{|c||X|}
    \hline
    Task & US Region Classification \\
    \hline\hline
    System Prompt & What US region is the following location in? Possible answers are ``SOUTH'', ``NON-SOUTH'', or ``N/A''. \\
    \hline
    Expected input & User location information \\
    \hline
    Default Label & N/A \\
    \hline
    Accuracy (\%) & 100.0 \\
    \hline
    \# Training Samples & 100 \\
    \hline
    \# Validation Samples & 100 \\
    \hline
    Data Source & Hand-crafted \\
    \hline
    Base Model & gpt-3.5-turbo-1106 \\
    \hline
    Trained Tokens & 13491 \\
    \hline
    \# Epochs & 3 \\
    \hline
    Final Training Loss & 0.0000 \\
    \hline
    \# Steps & 291 \\
    \hline
    Training Time & 00:13:18 \\
    \hline
  \end{tabularx}
\end{table*}

\begin{table*}[htp]
  \centering
  \caption{The system prompts, default classifications (in case the model produces an unexpected output), accuracies, and training/validation information of the fine-tuned GPT-3.5 model for the task of personal attack detection. Note that the system prompt was broken to fit it in the page. The actual prompt is single-line. \label{tab:pad}}
  \begin{tabularx}{\textwidth}{|c||X|}
    \hline
    Task & Personal Attack Detection \\
    \hline\hline
    System Prompt & Does the following text contain a personal attack? Possible answers are ``YES'' or ``NO''. \\
    \hline
    Expected input & User post \\
    \hline
    Default Label & NO \\
    \hline
    Accuracy (\%) & 96.0 \\
    \hline
    \# Training Samples & 2094 \\
    \hline
    \# Validation Samples & 2094 \\
    \hline
    Data Source & \citet{zhang-etal-2018-conversations} \\
    \hline
    Base Model & gpt-3.5-turbo-1106 \\
    \hline
    Trained Tokens & 816762 \\
    \hline
    \# Epochs & 3 \\
    \hline
    Final Training Loss & 0.0000 \\
    \hline
    \# Steps & 1501 \\
    \hline
    Training Time & 01:15:17 \\
    \hline
  \end{tabularx}
\end{table*}

\section{Aggression, Response, and Retaliation Rates}
\label{sec:fullmetrics}

In Table \ref{tab:fullmetrics}, we give the per-speaker and per-interaction aggression, response, and retaliation rates for every considered dataset. Note that due to the insufficient statistical power in \texttt{reddit-corpus-small}, \texttt{wikiconf}, and \texttt{wiki-corpus}, their numbers should not be used to draw any conclusions about the hypotheses~\cite{button2013power}.

\begin{table*}
  \caption{The per-speaker and per-interaction rates for each metric, US region, and dataset. The metric labels ``AGG,'' ``RESP,'' and ``RET'' represent aggression, response, and retaliation rates, respectively. For each rate, the sample size is equal to the number of people who posted, were personally attacked, and responded to a personal attack at least once, respectively. In this paper, we only analyze the \texttt{subreddit-changemyview} dataset since the others do not contain a sufficient number of interactions involving personal attacks. \label{tab:fullmetrics}}
  \centering
  \begin{tabular}{|c|c|c||c|c|c|c|}
    \hline
    \multirow{2}{*}{Dataset} & \multirow{2}{*}{Per} & \multirow{2}{*}{Metric} & \multicolumn{2}{c|}{Non-South} & \multicolumn{2}{c|}{South} \\
    \cline{4-7}
    & & & Rate (\%) & \# Samples & Rate (\%) & \# Samples \\
    \hline\hline
    \multirow{6}{*}{\small \textbf{\texttt{subreddit-changemyview}}} & \multirow{3}{*}{\textbf{Speaker}} & \textbf{AGG} & \textbf{9.1} & \textbf{4491} & \textbf{9.4} & \textbf{1674} \\
    \cline{3-7}
    & & \textbf{RESP} & \textbf{51.4} & \textbf{1318} & \textbf{51.4} & \textbf{453} \\
    \cline{3-7}
    & & \textbf{RET} & \textbf{20.6} & \textbf{849} & \textbf{25.3} & \textbf{292} \\
    \cline{2-7}
    & \multirow{3}{*}{\textbf{Interaction}} & \textbf{AGG} & \textbf{8.8} & \textbf{8507/96146} & \textbf{9.4} & \textbf{4820/51365} \\
    \cline{3-7}
    & & \textbf{RESP} & \textbf{54.5} & \textbf{4113/7552} & \textbf{56.3} & \textbf{2658/4717} \\
    \cline{3-7}
    & & \textbf{RET} & \textbf{23.0} & \textbf{947/4113} & \textbf{28.6} & \textbf{761/2658} \\
    \hline\hline
    \multirow{6}{*}{\small \texttt{reddit-corpus-small}} & \multirow{3}{*}{Speaker} & AGG & 16.5 & 2522 & 16.1 & 943 \\
    \cline{3-7}
    & & RESP & 52.1 & 528 & 46.7 & 171 \\
    \cline{3-7}
    & & RET & 26.0 & 211 & 15.4 & 58 \\
    \cline{2-7}
    & \multirow{3}{*}{Interaction} & AGG & 17.0 & 982/5780 & 14.7 & 277/1878 \\
    \cline{3-7}
    & & RESP & 34.4 & 342/993 & 31.8 & 90/283 \\
    \cline{3-7}
    & & RET & 29.2 & 100/342 & 15.6 & 14/90 \\
    \hline\hline
    \multirow{6}{*}{\texttt{wikiconv}} & \multirow{3}{*}{Speaker} & AGG & 5.5 & 80965 & 6.1 & 22486 \\
    \cline{3-7}
    & & RESP & 46.3 & 1365 & 43.1 & 350 \\
    \cline{3-7}
    & & RET & 18.2 & 142 & 22.7 & 37 \\
    \cline{2-7}
    & \multirow{3}{*}{Interaction} & AGG & 2.8 & {\small 15627/556156} & 2.2 & {\small 2793/125372} \\
    \cline{3-7}
    & & RESP & 14.2 & 271/1912 & 16.1 & 75/467 \\
    \cline{3-7}
    & & RET & 19.2 & 52/271 & 20.0 & 15/75 \\
    \hline\hline
    \multirow{6}{*}{\texttt{wiki-corpus}} & \multirow{3}{*}{Speaker} & AGG & 5.4 & 1429 & 7.5 & 400 \\
    \cline{3-7}
    & & RESP & 35.8 & 114 & 35.7 & 29 \\
    \cline{3-7}
    & & RET & 19.1 & 29 & 16.7 & 6 \\
    \cline{2-7}
    & \multirow{3}{*}{Interaction} & AGG & 3.0 & 287/9594 & 4.4 & 69/1578 \\
    \cline{3-7}
    & & RESP & 23.9 & 47/197 & 15.4 & 6/39 \\
    \cline{3-7}
    & & RET & 23.4 & 11/47 & 16.7 & 1/6 \\
    \hline
  \end{tabular}
\end{table*}

\section{R Code for Fitting a Mixed-Effect Model}
\label{sec:rcode}

The mixed-effect model for the per-speaker retaliation rates for the \texttt{subreddit-changemyview} dataset is fit using the \texttt{lme4} package in R.

\begin{verbatim} 
 glmer(retaliation ~ speaker_2_us_region +
   (1 | speaker_2_id),  family = "binomial")
\end{verbatim}

Here, the variables \texttt{speaker\_2\_us\_region} and \texttt{speaker\_2\_id} are the region and the id of a person who started an interaction, was personally attacked, and may potentially retaliate back.

\section{Statistics on the Ongoing Fights Detected by our System}
\label{sec:fullfp}

Up to 100 samples of user interactions in each dataset (including the ones that were not analyzed) flagged to involve personal attacks were analyzed by a human annotator to see if they truly constitute an ongoing online fight. As tabulated in Table \ref{tab:fullfp}, we found that, for the \texttt{subreddit-changemyview} dataset, the true ongoing fight rate was 86.0\%, high enough to ensure our analysis is correct. Note that the ongoing fight rates vary from 71.0\% (\texttt{reddit-corpus-small}, 75.0\% (\texttt{wiki-corpus}), to 91.0\% (\texttt{wikiconv}) for the other datasets.

\begin{table*}[!htp]
  \caption{The percentages of the sampled retaliatory interactions that a human annotator labeled as an ongoing fight. By showing that most samples we flagged as retaliatory to indeed be a part of an ongoing online fight, we validate the robustness of our analysis. \label{tab:fullfp}}
  \centering
  \begin{tabular}{|c||c|c|}
    \hline
    \multirow{2}{*}{Dataset} & \multicolumn{2}{c|}{Ongoing Fights} \\
    \cline{2-3}
    & Rate (\%) & \# Samples \\
    \hline\hline
    \textbf{\texttt{subreddit-changemyview}} & \textbf{86.0} & \textbf{86/100} \\
    \hline
    \texttt{reddit-corpus-small} & 71.0 & 71/100 \\
    \hline
    \texttt{wikiconv} & 91.0 & 61/67 \\
    \hline
    \texttt{wiki-corpus} & 75.0 & 9/12 \\
    \hline
  \end{tabular}
\end{table*}

\end{document}